\documentclass[aps,prd,reprint,showpacs,floatfix,nofootinbib,superscriptaddress]{revtex4-1}
\usepackage{amsmath}
\usepackage{amssymb}
\usepackage{latexsym}
\usepackage{graphics}
\usepackage{graphicx}
\usepackage{slashed}
\usepackage{color}
\usepackage{comment}

\usepackage{bm}
\usepackage[colorlinks=true,citecolor=cyan,urlcolor=blue,bookmarks=true,bookmarksopen=true,bookmarksnumbered=true,bookmarksopenlevel=3]{hyperref}

\definecolor{airforceblue}{rgb}{0.36, 0.54, 0.66}
\definecolor{steelblue}{rgb}{0.27, 0.51, 0.71}
\definecolor{amber}{rgb}{1.0, 0.49, 0.0}

\makeatletter
\newsavebox\myboxA
\newsavebox\myboxB
\newlength\mylenA

\newcommand*\xoverline[2][0.75]{%
    \sbox{\myboxA}{$\m@th#2$}%
    \setbox\myboxB\null
    \ht\myboxB=\ht\myboxA%
    \dp\myboxB=\dp\myboxA%
    \wd\myboxB=#1\wd\myboxA
    \sbox\myboxB{$\m@th\overline{\copy\myboxB}$}
    \setlength\mylenA{\the\wd\myboxA}
    \addtolength\mylenA{-\the\wd\myboxB}%
    \ifdim\wd\myboxB<\wd\myboxA%
       \rlap{\hskip 0.5\mylenA\usebox\myboxB}{\usebox\myboxA}%
    \else
        \hskip -0.5\mylenA\rlap{\usebox\myboxA}{\hskip 0.5\mylenA\usebox\myboxB}%
    \fi}
\makeatother

\begin{document}

\title{Like-sign dileptons with mirror type composite neutrinos at the HL-LHC}

\author{\textsc{M.~Presilla}}\email[{\bf Corresponding Author,}\\ Email: ]{matteo.presilla@cern.ch} 
\affiliation{Dipartimento di Fisica e Astronomia "Galileo Galielei", Universit\`{a} degli Studi di Padova, Via Marzolo, I-35131, Padova, Italy}
\affiliation{Istituto Nazionale di Fisica Nucleare, Sezione di Padova, Via Marzolo, I-35131, Padova, Italy}

\author{\textsc{R.~Leonardi}}
\affiliation{Istituto Nazionale di Fisica Nucleare, Sezione di Perugia, Via A.~Pascoli, I-06123 Perugia, Italy}

\author{\textsc{O.~Panella}}
\affiliation{Istituto Nazionale di Fisica Nucleare, Sezione di Perugia, Via A.~Pascoli, I-06123 Perugia, Italy}

\date{\today}

\begin{abstract}
Within  a mirror type assignment for the excited composite fermions the neutrino mass term is  built up from a Dirac mass, $m_*$,  which gives  the mass of charged lepton component of the $SU(2)$, right-handed, doublet, and a Majorana mass, $m_L$, for the left-handed component (singlet) of the excited neutrino.
The mass matrix is diagonalized leading to two Majorana mass eigenstates. The active neutrino field $\nu^*_R$ is thus a superposition of the two mass eigenstates with mixing coefficients which depend on the ratio $m_L/m_*$.
We discuss the prospects of discovery of these physical states at the HL-LHC as compared with the previous searches of composite Majorana neutrinos at the LHC based on sequential type Majorana neutrinos.
\end{abstract}


\maketitle

\textit{Introduction -} A composite scenario~\cite{Eichten:1983hw,Peskin:1985symp,Cabibbo:1983bk,Baur:1989kv,Terazawa:1976xx,Terazawa:1979pj,Terazawa:2011ci,Terazawa:2014dla,Terazawa:2015bsa,Meyer:1984wfa,Bogolyubov:2011sua}, where at a sufficiently high energy  scale $\Lambda$ (compositeness scale) the standard model leptons and quarks show the effects of an internal substructure, has triggered  considerable recent interest both from the theoretical~\cite{Panella:2017spx,Biondini:2012ny,Leonardi:2014aa,Biondini:2017fut} and experimental~\cite{Aad:2013ab,Khachatryan:2016ac} point of view. In particular, recent studies~\cite{Leonardi:2015qna} have concentrated in searching for heavy composite Majorana neutrinos at the LHC. 
A recent CMS study has searched for a  heavy composite Majorana neutrino ($N$), using 2.6 fb$^{-1}$ data of the 2015 Run II  at $\sqrt{s}=13$ TeV~\cite{CMS-PAS-EXO-16-026,Sirunyan:2017xnz}. Heavy composite neutrino masses are excluded, at 95\% CL, up to $m_N=4.60$ TeV in the $eeqq$ channel and $m_N=4.70$ TeV in the $\mu\mu qq$ channel for a value of $\Lambda=m_N$.

We discuss here a variant of the model analyzed in~\cite{Leonardi:2015qna,Sirunyan:2017xnz} taking up the scenario in which the excited fermions are organized with a mirror SU(2) structure relative to the SM fermions, i.e. the right-handed components form an SU(2) doublet while the left-handed components are singlets~\cite{Olive:2014aa}. We construct a general Dirac-Majorana mass term and discuss its mass spectrum along with prospects of observing the resulting lepton number violating signatures at the High-Luminosity Large Hadron Collider (HL-LHC)~\cite{Apollinari:2015bam}.
The HL-LHC phase intends to push the performance of the LHC increasing the potential for discoveries by improving the luminosity by a factor of 10 and the Centre-of-Mass energy by 1 TeV, bringing them to 3 $ab^{-1}$ and 14 TeV, respectively. Hence, the exploration of very rare and yet-unobserved deviations to the Standard Model could be feasible thanks to this very high expected statistics.
\vspace{1em}

\textit{Mirror type Model -} In analogy with the usual procedure adopted in see-saw type extensions of the standard model, we may give a (lepton number violating) Majorana mass term  to the left-handed excited neutrino ($\nu_L^*$), which is a singlet and does not actively participate to the gauge interactions in Eq.~\ref{mirror} --\emph{sterile neutrino}--, while a (lepton number conserving) Dirac mass term $m^*$ is associated to the right-handed component $\nu_R^*$  which belongs to the $SU(2)$ doublet and does participate in the gauge interactions -- \emph{active neutrino}.
We can thus write down the excited neutrino Dirac-Majorana mass term appearing in the Lagrangian density of the model as:
\begin{equation}\label{mass_term1}{\cal L}^{\text{D+M}} = -\frac{1}{2} m_L \,\bar{\nu}^*_L ({\nu}^*_L)^c
- m_*\, \bar{\nu}^*_L \nu^*_R\,  +\text{h.c.}\,.  
\end{equation} 

The Lagrangian mass term is easily diagonalized following standard procedures, obtaining two Majorana mass eigenstates, $\nu_{1,2}$, with (positive) mass eigenvalues given by: 
\begin{equation}
\label{masseigenstates}
m_{1,2}=\sqrt{m_*^2 +\left(\frac{m_L}{2}\right)^2} \mp \frac{m_L}{2}
\end{equation}

Interacting states can be written as a mixing of the mass eigenstate according to the relation:
\begin{subequations}
\label{inverted2}
\begin{align}
\label{inverted2a}
(\nu_{L}^*)^c &= \,- i\cos\theta\, \nu_{1R} + \sin\theta \,\nu_{2R}\\ 
\label{inverted2b}\nu_{R}^* &=\, \phantom{-}i\sin\theta\, \nu_{1R} + \cos\theta \,\nu_{2R}
\end{align}
\end{subequations} 
with the mixing angle $\theta$ written in terms of the two masses of Dirac and Majorana:
\begin{equation}
\label{mixingangle}
\theta = -\frac{1}{2} \text{arctg}\left( \frac{2m_*}{m_L}\right).
\end{equation}

\begin{figure*}[t!]
\includegraphics[width=0.48\linewidth,height=6.0cm]{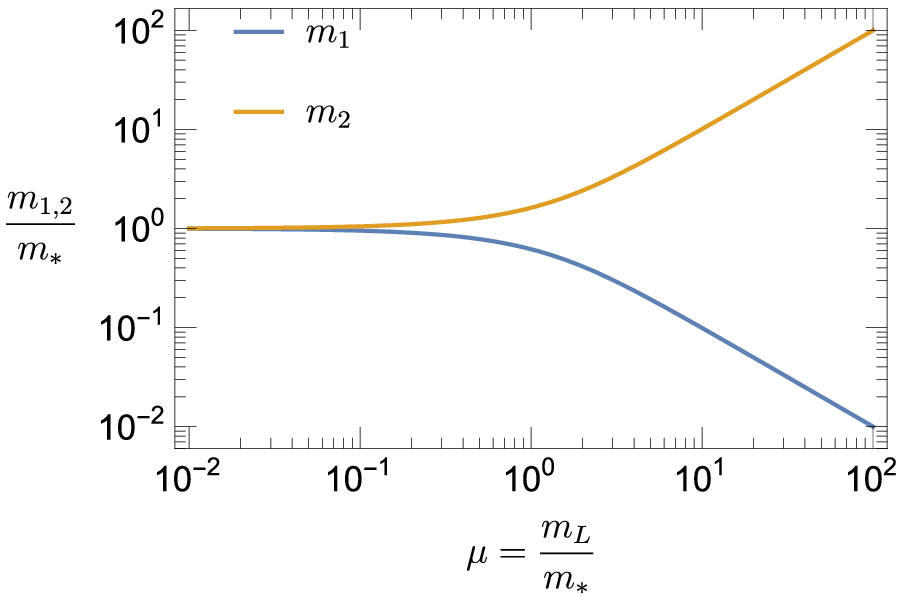}\hspace{0.5cm}
\includegraphics[width=0.48\linewidth,height=6.0cm]{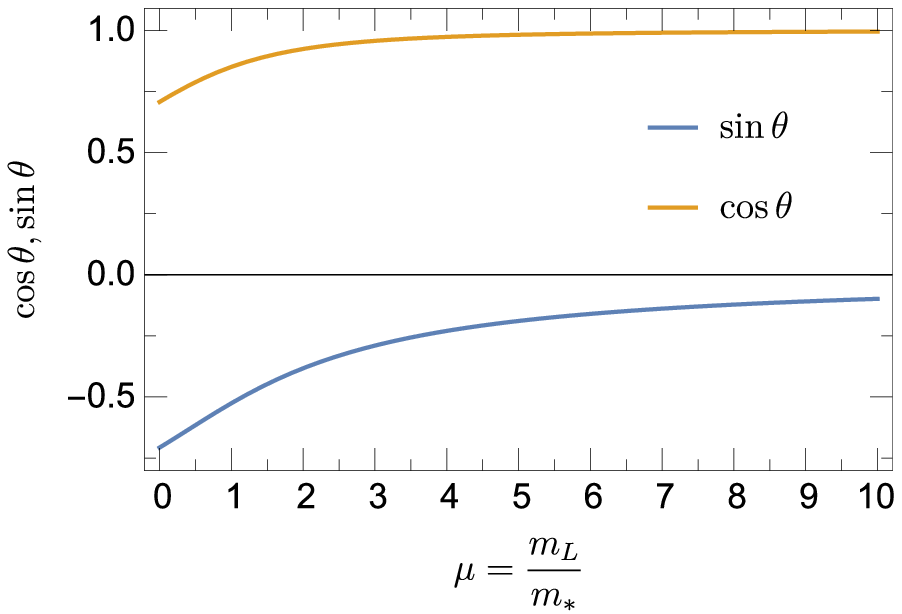}
\caption{(Left-panel) Mass eigenstates $m_{1,2}$ (in units of $m_*$) from Eq.~\protect\eqref{masseigenstates} as a function of the parameter $\mu=m_L/m_*$. (Right-panel) Mixing coefficients $\cos\theta$ and $\sin\theta$ from Eq.~\protect\eqref{mixingangle} as a function of the parameter $\mu=m_L/m_*$}
\end{figure*}

Now we take into account all the relevant effective couplings of these particles. In the usual mirror type model, we assume that the excited neutrino and the excited electron are grouped into left-handed singlets and  a right-handed $SU(2)$ doublet.
The corresponding gauge mediated Lagrangian between the left-handed SM doublet and the right-handed excited doublet via the $SU(2)_L\times U(1)_Y$ gauge fields~\cite{Takasugi:1995bb,Olive:2014aa}, which is of the magnetic type for current conservation, can be written down:
\begin{equation}
\label{mirror}
{\cal L} =\frac{1}{2 \Lambda} \, \bar{L}_R^* \sigma^{\mu\nu}\left( gf \frac{\bm{\tau}}{2}\cdot \bm{W}_{\mu\nu} +g'f' Y B_{\mu\nu}\right) L_L +h.c. \, ,
\end{equation}
where $L^T =({\nu_\ell}_L, \ell_L)$ is the ordinary $SU(2)_L$ lepton doublet, $g$ and $g'$ are the $SU(2)_L$ and $U(1)_Y$ gauge couplings and $\bm{W}_{\mu\nu}$, $B_{\mu\nu}$ are the field strength for the $SU(2)_L$ and $U(1)_Y$ gauge fields; $f$ and $f'$ are dimensionless couplings usually set equal to unity.
The relevant charged current (gauge) interaction of the excited (active) Majorana neutrino $\nu_R^*$ is easily derived from Eq.~\ref{mirror}:
\begin{equation}
\label{lagGI}
\mathcal{L}_{\text{G}}=\frac{gf}{\sqrt{2}\Lambda}\, \xoverline{\nu}^*_R \, \sigma^{\mu \lambda} \, \ell_L \,  \partial_{\mu}  W_{\lambda} \,  + h. c.
\end{equation}
and can be written out explicitly in terms of the Majorana mass eigenstates through Eq.~\eqref{inverted2b}:
\begin{equation}
\label{lagGImass}
\mathcal{L}_{\text{G}}=\frac{gf}{\sqrt{2}\Lambda}\,\left( -i\sin\theta\, \xoverline{\nu}_{1} +\cos\theta\, \xoverline{\nu}_{2}\right) \, \sigma^{\mu \lambda} \, \ell_L \,\,  \partial_{\mu} \, W_{\lambda} \,  + h. c.\, .
\end{equation}

Contact interactions between ordinary and excited fermions may arise by constituent exchange if the fermions have common constituents, and/or by the exchange of the binding quanta of the new unknown interaction, whenever such binding quanta couple to the constituents of both particles~\cite{Peskin:1985symp,Olive:2014aa}. The dominant effect is expected to be given by the 6-dimension four-fermion  interactions which scale with the inverse square of the compositeness scale $\Lambda$: 
\begin{subequations}
\label{contact}
\begin{align}
\label{Lcontact}
\mathcal{L}_{\text{CI}}&=\frac{g_\ast^2}{\Lambda^2}\frac{1}{2}j^\mu j_\mu ,\\
\label{Jcontact}
j_\mu&=\eta_L\bar{f_L}\gamma_\mu f_L+\eta\prime_L\bar{f^\ast_L}\gamma_\mu f^\ast_L+\eta\prime\prime_L\bar{f^\ast_L}\gamma_\mu f_L + h.c.\nonumber\\&\phantom{=} +(L\rightarrow R)
\end{align}
\end{subequations}
where $g_*^2 = 4\pi$ and the $\eta$ factors are usually set equal to unity. In this work the right-handed currents will be neglected for simplicity.

The single production $q\bar{q}' \to \nu^*\ell$ proceeds through flavour conserving but non-diagonal terms, in particular with currents like  the third term in Eq.~\ref{Jcontact} which couple excited states with ordinary fermions:
\begin{equation}
\label{lagCI}
\mathcal{L}_{\text{CI}}= \frac{g_\ast^2}{\Lambda^2} \, \bar{q}_L\gamma^\mu  q_L' \, \xoverline{\nu^*_L}\gamma_\mu \ell_L \, .
\end{equation} 
The contact interactions in Eq.~\eqref{lagCI} can be written out explicitly in terms of the Majorana mass eigenstates $\nu_i$ using Eq.~\eqref{inverted2b}: 
\begin{equation}
\label{lagCImass}
\mathcal{L}_{\text{CI}}= \frac{g_\ast^2}{\Lambda^2} \, \bar{q}_L\gamma^\mu  q_L' \, \left( - i\cos\theta\, \xoverline{\nu}_{1} + \sin\theta \,\xoverline{\nu}_{2} \right)\,\gamma_\mu \ell_L \, .\\
\end{equation}

\begin{figure*}[t!]
\includegraphics[width=0.48\linewidth, height=6. cm]{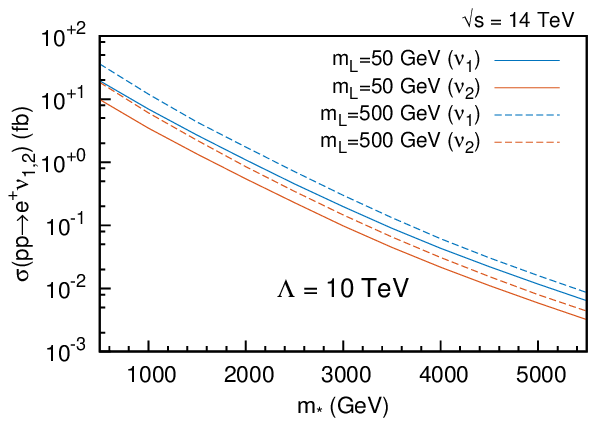}
\includegraphics[width=0.48\linewidth, height=6. cm]{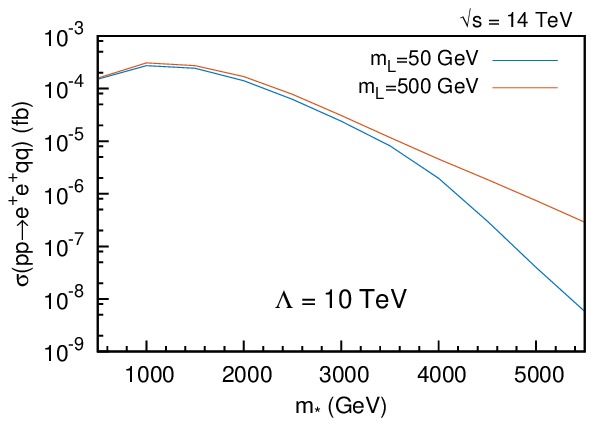}
\caption{(Left-panel) Production cross section, at $\sqrt{s}=14$ TeV, for the two mass eigenstates $pp\rightarrow e^{+}\nu_{1,2}$ for two different Majorana mass values, $m_L=50, 500$ GeV. (Right-panel) Cross section for the like-sign dileptons signature, $pp \to e^{+}e^{+} qq$ at the HL-LHC ($\sqrt{s}=14$ TeV).}\label{xsec}
\end{figure*}


The model needs to be implemented in a Monte Carlo generator to extract predictions. Here we have obtained the Feynman rules of the model thanks to the Mathematica package FeynRules~\cite{Christensen:2008py}. The MC simulation samples and the numerical computations are then mostly obtained with Madgraph~\cite{Alwall:2014hca}, complementing the results with a validation obtained with the tree-level simulator CalcHEP~\cite{Belyaev20131729}. 

\vspace{1em}

\begin{figure}[htpb]
\includegraphics[width=.4\textwidth]
{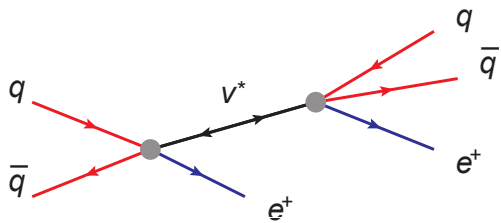}
\caption{The like-sign dileptons and dijets signature mediated by the heavy composite Majorana neutrino.}\label{fig:FeynDiag}
\end{figure}

\begin{figure}[t!]
\includegraphics[width=.49\textwidth]
{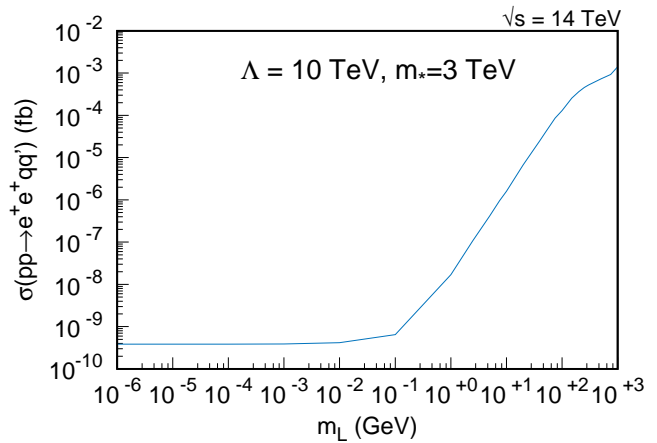}
\caption{The like-sign dileptons and dijets cross section as a function of $m_L$, given $\Lambda = 10$ TeV and $m_* = 3$ TeV .}\label{mL0}
\end{figure}

\textit{Probe at the HL-LHC -}
In the compositeness framework discussed above, the heavy Majorana neutrino could trigger processes with Lepton Number Violation (LNV). 
In particular, the production of a heavy neutrino $\nu^*$ in association with a charged lepton could be followed by the subsequent decay of the neutrino to a like-sign lepton plus two jets coming from the hadronization of the quarks in the detector. It is important to emphasize that, in this version of the model, the mediator acts as a mixture of the two mass eigenstates.
The whole process results in the like-sign dileptons and dijets signature, that is the golden channel for LNV searches in hadron collider experiments.   
This is shown diagrammatically in Fig.~\ref{fig:FeynDiag}. 
Here we focus on the two-positrons final state, because of the higher-luminosity of the partons involved in this channel in proton-proton collisions with respect to the negatively charged case.
Fig.\ref{xsec} shows the behaviour of the production cross section for the two mass eigenstates and for the full LNV process for different model scenarios.
It is worth to notice that in the limit $m_{L} \rightarrow 0$ the cross-section of the like-sign dileptons and dijets process goes to zero, as the process becomes essentially mediated by Dirac neutrinos. This is confirmed by the behaviour of the signal cross section with respect to the parameter $m_L$ as shown in Fig.~\ref{mL0}. \\

In order to study the potentiality of a successful detection of this kind of particle, LHE samples coming from the MC generators are interfaced with the Fast-Simulation framework Delphes~\cite{deFavereau:2013fsa}, considering a CMS Phase-2~\cite{CMSCollaboration:2015zni} parametrization without considering any pileup effect. 
Hence, the potential for discovery at HL-LHC in the three-dimensional parameter space $(\Lambda, m_{L}, m_{*})$ can briefly discussed.

\begin{figure}[t!]
\includegraphics[scale=1.35]
{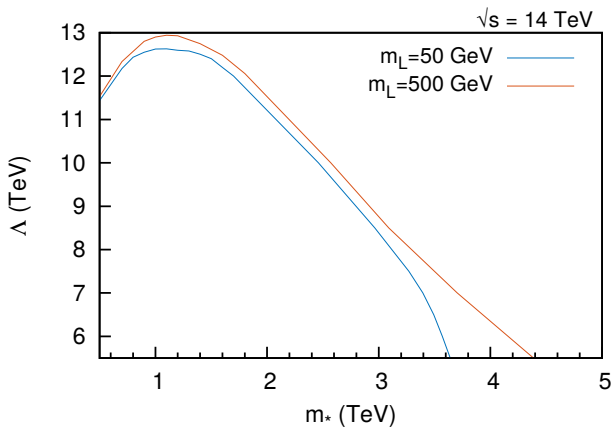}
\caption{5-$\sigma$ level contour curves of statistical significance for two hypothesis of $m_L$, in the parameter space $(\Lambda,m_{*}) $ for $\sqrt{s} = 14 $ TeV and for an integrated luminosity $L = 3000$ fb$^{-1}$. The region below a curve allows a significance greater than 5-$\sigma$ in case of a future discovery.}\label{contour}
\end{figure}

Standard Model processes that could mimic the detection of a LNV signal in this rather clean signature are mainly the triple W boson production, $pp \rightarrow W^+ W^+ W^-$, and the top quark pair production $pp \rightarrow t \bar{t}$, the former being the dominant background source. 
Since the kinematic features of the final state reconstructed objects are similar to those of Ref.~\cite{Leonardi:2015qna}, we lower the background contribution by imposing two cuts on the leading lepton ($p_T (e_1) > 110$ GeV) and on the second-leading lepton ($p_T (e_1) > 35$ GeV).  
This particular signal region allows an efficiency in selecting signal around the $80 \%$, while beating the background sources with efficiency of $0.00044 \%$ for the $t\bar{t}$ and of $0.0034 \%$ for the  $W^+ W^+ W^-$. \\
The statistical significance is defined by the relation 
\begin{equation}
    S=\dfrac{\mathcal{L} \sigma_{\text{sig}} \epsilon_{\text{sig}}}{\sqrt{\mathcal{L} \sigma_{\text{bkg}} \epsilon_{\text{bkg}}}},
\end{equation}
where $\epsilon_{\text{sig}},\epsilon_{\text{bkg}}$ are respectively the cumulative efficiencies of the signal and the background due to the selection.\\
Figure~\ref{contour} shows the 5-$\sigma$ contours in the $(\Lambda, m_*)$ for two benchmark choices of the lepton number violating parameter $m_{L}=50, 500$ GeV. 

\begin{figure}[t!]
\includegraphics[scale=1.35]
{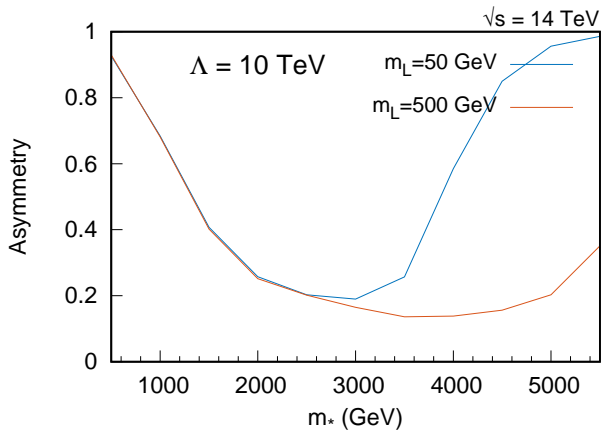}
\caption{Asymmetry in the production of like-sign signature and opposite-sign signature prompted by a heavy composite Majorana neutrino $\nu^*$.}\label{asymmetry}
\end{figure}

Another interesting feature of this version of the model is that the presence of two heavy neutrino mass eigenstates, which have also opposite $\mathcal{CP}$ eigenvalues, can lead to an asymmetry in the production rate of like-sign dileptons signal and opposite-sign signal~\cite{Dev:2015aa}. In particular, a statistically significant nonzero ratio of same- and opposite-sign dileptons signal events could be used to test the relative strength between the Dirac and Majorana nature of the
heavy neutrinos at the HL-LHC. 

We have studied the charge asymmetry 
\begin{equation}
\label{asymmetrydef}
{\cal A}=\frac{\sigma_{e^+ e^- j j}-\sigma_{e^+ e^+ jj}}{\sigma_{e^+ e^- j j}+\sigma_{e^+ e^+ j j}},    
\end{equation}
for a fixed value of the scale $\Lambda$. We expect that the compositeness scale does not play a crucial role in the asymmetry because  when taking the ratio in Eq.~\ref{asymmetrydef} it will essentially cancel out. In addition it should also not play a significant role in the yields of the like sign versus the opposite sign. 
As shown in Fig.~\ref{asymmetry}, the asymmetry depends both on the Majorana mass $m_L$ and on the Dirac mass $m_*$ without reaching the $\mathcal{A}=0$, that would correspond to a process mediated by a single Majorana neutrino. 

As the Dirac mass reaches higher values, we see that the like-sign channel is suppressed ($\mathcal{A}=1$), being the magnitude of the suppression dependent on the parameter $m_L$.

\vspace{1em}

\textit{Conclusions -} We have presented a mechanism of lepton  number violation within a mirror type model compositeness scenario. The mirror type case is realized when the left components of the excited states are singlets while the right components are active doublets which do participate to gauge transition interactions with SM fermions. We therefore introduce a \emph{left}-handed Majorana neutrino singlet of mass $m_L$ while the right-handed component belongs to a doublet and has a Dirac mass $m_*$.  This situation is exactly specular to the one encountered in typical see-saw models where the sterile neutrino is the right-handed component and the active one is the left-handed component.
Diagonalization of the mass matrix gives two Majorana mass eigenstates whose phenomenology at the HL-LHC is presented. We  provide the 5-$\sigma$ contour curves of the statistical significance, for two different values of the Mjorana mass $m_L=50,500$ GeV, of the like-sign dileptons and dijet signature  giving  indications about the discovery potential at the CMS Phase-2 detector.
We have additionally shown that the charge asymmetry  is a potential observable that could allow to distinguish between different values of $m_L$ in the regime of high values of the Dirac mass ($m_* \in [ 2500, 5000]$ GeV for $\Lambda=10$ TeV). 

\begin{acknowledgments}
The authors acknowledge constant and encouraging support from the CMS group,  University of Perugia, Department of Physics and Geology and INFN, Sezione di Perugia.
\end{acknowledgments}

\bibliographystyle{unsrt}

\end{document}